\documentstyle[12pt]{article}
\def\UV{ultraviolet }
\def\i{{\rm i}}
\def\F{{\cal F}}
\def\G{{\cal G}}
\def\H{{\cal H}}
\def\C{{\cal C}}
\def\D{{\cal D}}
\def\E{{\cal E}}

\def\R{{\cal R}}
\def\S{{\cal S}}
\def\to{\rightarrow}
\newcommand{\lra}{\leftrightarrow}
\newcommand{\la}{\langle}
\newcommand{\ra}{\rangle}
\def\vspaceinarray{\nonumber ~&~&~\\}
\newcommand{\Nc}{N_c}
\newcommand{\Nf}{N_f}
\def\A#1#2{\la#1#2\ra}
\def\B#1#2{[#1#2]}
\def\s#1#2{s_{#1#2}}
\def\P#1#2{{\cal P}_{#1#2}}
\def\L#1#2{\left\{#2\right\}_{#1}}

\def\Li{{\rm Li_2}}

\newcommand{\ve}{\varepsilon}
\newcommand{\beq}{\begin{equation}}
\newcommand{\eeq}{\end{equation}}
\newcommand{\beqn}{\begin{eqnarray}}
\newcommand{\eeqn}{\end{eqnarray}}
\newcommand{\beqns}{\begin{eqnarray*}}
\newcommand{\eeqns}{\end{eqnarray*}}

\def\Am{{\cal A}}
\def\nn{\nonumber}

\def\mus#1#2{\left(-\frac{\mu^2}{s_{#1#2}}\right)^\varepsilon}
\def\qb{\bar{q}}
\def\Qb{\bar{Q}}

\begin{document}

\begin{titlepage}
\vspace*{-2cm}
\begin{flushright}
ETH-TH/94-14\\
KLTE-DTP/94-3\\
May 25, 1994 \\
\end{flushright}
\vskip .5in
\begin{center}
{\Large\bf
One-loop radiative corrections to the helicity amplitudes of QCD
processes
involving four quarks and one gluon
}\footnote{
Work supported in part by the Swiss National Science Foundation,
the Hungarian National Science Foundation and the
Universitas Foundation of the Hungarian Commercial Bank}\\
\vskip 1cm
{\large Zoltan Kunszt$^a$, Adrian Signer$^a$ and Zolt\'an
  Tr\'ocs\'anyi$^{a,b}$} \\
\vskip 0.2cm
$^a$ Theoretical Physics, ETH, \\
Z\"urich, Switzerland  \\
\vskip 0.2cm
$^b$ Department of Theoretical Physics, KLTE, \\
Debrecen, Hungary \\
\vskip 1cm
\end{center}

\begin{abstract}
\noindent
We present analytic results for the one-loop corrections of the
helicity amplitudes of the QCD five-parton subprocesses involving
four quarks and one gluon obtained with a standard Feynman diagram
calculation using dimensional reduction.
\end{abstract}
\end{titlepage}
\setcounter{footnote}{0}

\bigskip

The technical developments achieved recently in perturbative QCD calculations
(helicity method \cite{Xu87,Gun85,KleissWJS85,Man91}, string theory based
derivations \cite{Ber92,Ber93}) made
calculations of  one-loop  corrections to helicity amplitudes
up to five-parton  and perhaps also to six-parton processes feasible.
As a first result,  Bern, Dixon and Kosower published recently
the one-loop corrections to the five-gluon amplitude in QCD and
in N=1,2,4 supersymmetric Yang-Mills theories.
They used string based technique  and helicity method and obtained
remarkably short analytic answer.
In a previous paper \cite{KST94} we presented simple analytic result for
the singular parts of all helicity amplitudes of all five-parton
processes (in full agreement with the results of \cite{Ber93}
in the case of the five gluon amplitudes).
In this letter we present the complete one-loop corrections
for processes involving four quarks (equal or unequal flavors) and one gluon.
We used conventional Feynman-diagram method, however,  the use of
the  helicity technique was decisively important.
Our method was tested previously by the diagrammatic evaluation of
the  one-loop corrections of  the  helicity amplitudes
of all $2\to 2$ parton processes \cite{ZKASZT93}. A systematic
description together with the non-trivial details of our
calculation will be presented in a later publication.
Here we mention only that the tensor integrals have been reduced to
scalar integrals with a reduction method similar to \cite{OldVer90}.
Calculating the integrals this way and the use of the helicity method
allow us to eliminate all integrals which are more complicated than
pentagon (box) tensor-integrals with one (two) integration momenta in
the numerator. As a result, the  gram determinants in the
denominator which blow up the size of intermediate expressions
were eliminated at the very beginning of the calculation.
For the sake of simplicity we performed the calculation in the
dimensional reduction scheme (in $D=4-2 \ve$ dimensions).
Transition rules to different schemes
--- such as conventional dimensional regularization, 't~Hooft--Veltman
--- have been derived in \cite{ZKASZT93}.

It is convenient to give our result  in a crossing symmetric form
for the unphysical channel where all particles are outgoing
$0 \to \qb \Qb Q q g$. The momenta of the partons are labeled as
\begin{equation}
0 \to {\rm antiquark_1}(\qb) + {\rm antiquark_2}(\Qb)
+ {\rm quark_2}(Q) + {\rm quark_1}(q) + {\rm gluon}(g)\ .
\label{label4q1g}
\end{equation}

\noindent
The color structure of the amplitudes is the same at one loop as at tree
level:

\begin{eqnarray}
\lefteqn{\Am^{(i)}(\qb,h_{\qb};\Qb,h_{\Qb};Q,h_Q;q,h_q;g,h_g) = } \\ \nn
&&g^3\left({g\over 4\pi}\right)^{2i}\left[\sum_{(q_1\ne q_2)\in \{q,Q\}}
(T^{g})_{q_1\qb_2}\delta_{q_2\qb_1}
a^{(i)}_{q_1\qb_2}(\qb,h_{\qb};\Qb,h_{\Qb};Q,h_Q;q,h_q;g,h_g)\right. \\ \nn
&&\qquad \quad-\left.\sum_{(q_1\ne q_2)\in \{q,Q\}}
\frac{1}{N_c}(T^{g})_{q_1\qb_1}\delta_{q_2\qb_2}
a^{(i)}_{q_1\qb_1}(\qb,h_{\qb};\Qb,h_{\Qb};Q,h_Q;q,h_q;g,h_g)\right],
\end{eqnarray}
where $i=0$ means tree level and $i=1$ means one-loop approximation.

At a given order in perturbation theory, there are only two
independent color subamplitudes because we have the symmetry properties
\beqn
a^{(i)}_{ q \Qb}(\qb,h_{\qb};\Qb,h_{\Qb};Q,h_Q;q,h_q;g,h_g) &=&
a^{(i)}_{ Q \qb}(\Qb,h_{\Qb};\qb,h_{\qb};q,h_q;Q,h_Q;g,h_g), \\
a^{(i)}_{ q \qb}(\qb,h_{\qb};\Qb,h_{\Qb};Q,h_Q;q,h_q;g,h_g) &=&
a^{(i)}_{ Q \Qb}(\Qb,h_{\Qb};\qb,h_{\qb};q,h_q;Q,h_Q;g,h_g).
\eeqn

Furthermore, we should consider only four helicity configurations.
If we change the sign of all helicities, we obtain the corresponding
amplitudes simply by replacing the spinor products $\la ...\ra\to -[...]$,
where the angle bracket and squared bracket denote spinor products with
minus-plus and plus-minus helicities,
 $$ \la pq \ra =\overline{\psi_-(p)}\psi_+(q) \quad {\rm and}\quad
 [pq]=\overline{\psi_+(p)}\psi_-(q) . $$
Due to helicity conservation along a fermion line, we have
$h_q=-h_{\qb}$ and $h_Q=-h_{\Qb}$. We present our result for positive
gluon helicity and label the remaining helicities with $h_q$ and $h_Q$.

In order to be able to write down the result for arbitrary values
of $h_q$ and $h_Q$, we introduce the helicity dependent momenta
\beqn
r(h_q)&=&q\quad {\rm if}\quad h_q=- \quad {\rm and}\quad
 r(h_q)=\qb \quad{\rm if}\quad h_q=+;
\\ \nn
R(h_Q)&=&Q\quad {\rm if}\quad h_Q=- \quad {\rm and}\quad
 R(h_Q)=\Qb \quad{\rm if}\quad h_Q=+.
\eeqn
We shall suppress the helicity dependence of $r$ and $R$.

For the sake of completeness, we recall the tree-level amplitudes
$a^{(0)}_{ij}$
\beq
a^{(0)}_{ij}(h_q,h_Q,+) = \i p_a(h_q,h_Q,+)\,\frac{\A ij}{\A ig \A gj}.
\eeq
We absorbed  the helicity dependence  completely  in the factor $p_a$
which is given by
\beq
p_a(h_q,h_Q,+) = (-1)^{\delta_{h_q h_Q}}\frac{\A rR^2}{\A q\qb \A Q\Qb},
\eeq
where $\delta$ is the usual Kronecker $\delta$.

At one loop, the result can  naturally be decomposed into soft contributions
$S_{ij}(h_q,h_Q,+)$ given in \cite{KST94}, \UV renormalization terms
$\R_{ij}(h_q,h_Q,+)$, terms coming from the expansion of collinear
$1/\epsilon$ singularities  $\C_{ij}(h_q,h_Q,+)$ given by vertex
and self energy integrals, finite terms  composed from spinor products,
dot products and single logarithms $\D_{ij}(h_q,h_Q,+)$,
and  finite terms $\E_{ij}(h_q,h_Q,+)$ containing  factors of $\Li$,
$\ln^2$  and $\pi^2$ coming from pentagon and box integrals.
The labels $i,j$ run over $q,\,Q,\,\qb,\,\Qb$ similarly to the
corresponding labels in the color subamplitudes $a_{ij}^{(1)}$.
The color subamplitude  $ a_{ij}^{(1)} $ has then the decomposition
\beqn
\lefteqn{a_{ij}^{(1)}(\qb,-h_q;\Qb,-h_Q;  Q,h_Q; q,h_q; g,+) = } \\
&&  {\cal S}_{ij}(h_q,h_Q,+) +  \R_{ij}( h_q, h_Q,+) +
 {\cal C}_{ij}( h_q, h_Q,+) + {\cal D}_{ij}( h_q, h_Q,+) +
    {\cal E}_{ij}( h_q, h_Q,+), \nn
\eeqn
where we suppressed the dependence on the  momenta $\qb,q,\Qb,Q,g $.
For completeness we recall  the soft contributions
\begin{eqnarray}
S_{ Q \qb}(h_q,h_Q,+) &=&
- \frac{c_\Gamma}{\ve^2} \left\{
\Nc a^{(0)}_{ Q \qb}(h_q,h_Q,+)
\left[\P \qb g + \P \Qb q + \P Q g \right] \right. \\ \nn
&& \qquad -\frac{1}{\Nc} a^{(0)}_{ Q \qb}(h_q,h_Q,+)
\left[-\P \qb \Qb + \P \qb Q + \P \qb q + \P \Qb Q + \P \Qb q - \P Q q
\right] \\ \nn
&&  \qquad -\frac{1}{\Nc} a^{(0)}_{ Q \Qb}(h_q,h_Q,+)
\left[-\P \qb \Qb + \P \qb g  + \P \Qb q - \P q g \right] \\ \nn
&&  \qquad -\frac{1}{\Nc} a^{(0)}_{ q \qb}(h_q,h_Q,+)
\left.\left[\P \Qb q - \P \Qb g
 - \P Q q + \P Q g \right]
\right\}
\end{eqnarray}
and
\begin{eqnarray}
S_{ Q \Qb}(h_q,h_Q,+) &=&
- \frac{c_\Gamma}{\ve^2} \left\{
\Nc a^{(0)}_{ Q \Qb}(h_q,h_Q,+)
\left[\P \qb q + \P \Qb g + \P Q g \right] \right. \\ \nn
&& \qquad -\frac{1}{\Nc} a^{(0)}_{ Q \Qb}(h_q,h_Q,+)
\left[-\P \qb \Qb + \P \qb Q + \P \qb q  + \P \Qb Q + \P \Qb q - \P Q q
\right]
 \\ \nn
&& \qquad +\Nc
 a^{(0)}_{ Q \qb}(h_q,h_Q,+)
\left[\P \qb \Qb - \P \qb q - \P \Qb g + \P q g \right] \\ \nn
&& \qquad +\Nc a^{(0)}_{ q \Qb}(h_q,h_Q,+)
\left.\left[-\P \qb q + \P \qb g + \P Q q - \P Q g \right] \right\},
\end{eqnarray}
where we introduced
$$
\P i j = \mus i j  \quad c_\Gamma={(4\pi)^{\ve}}
\frac{\Gamma^2(1-\ve)\Gamma(1+\ve)}{\Gamma(1-2\ve)} \ .
$$
We note that the helicity dependence of the soft contribution
is given by the Born factors $a^{(0)}_{ij}$. This holds for the
\UV renormalization contributions and the collinear $(1/\ve)$
singularity as well. We record the result in such a form that
the helicity dependence of the $\C$ terms is again
completely factored into the Born terms\footnote{Note that
the $\C$ terms are not uniquely determined since we can always shift
finite contributions between $\D$ and $\C$ terms.}.
\beqn
\R_{ij}(h_q,h_Q,+) &=&-{{(4\pi)^{\ve}}
\over {2\ \ve \Gamma(1-\ve)}}
( 11 \Nc - 2 \Nf)  \  a_{ij}^{(0)}(h_q,h_Q,+) \\
\vspaceinarray
\C_{Q \Qb} (h_q,h_Q,+) &=&
\frac{c_\Gamma}{\ve} a_{Q \Qb}^{(0)} (h_q,h_Q,+)
\left( \frac{2}{3}(\Nc - \Nf ) \P \qb q +
\frac{3}{2}\frac{1}{N_c} (\P \qb q + \P \Qb Q) \right) \\
&+& a^{(0)}_{Q \Qb} (h_q,h_Q,+) \left( \frac{29}{18} \Nc
+ \frac{13}{2} \frac{1}{\Nc} -  \frac{10}{9} \Nf \right) \nn \\
\vspaceinarray
\label{CQqb}
\C_{Q \qb} (h_q,h_Q,+) &=&
\frac{c_\Gamma}{\ve} a^{(0)}_{Q \qb} (h_q,h_Q,+)
\left( \frac{2}{3} (\Nc - \Nf ) \P \Qb Q +
\frac{3}{2} \frac{1}{N_c} (\P \qb q + \P \Qb Q) \right) \\
&+& a^{(0)}_{Q \qb} (h_q,h_Q,+) \left( \frac{29}{18} \Nc
+ \frac{13}{2} \frac{1}{\Nc} -  \frac{10}{9} \Nf \right)   \nn
\eeqn

As expected, the next-to-leading order corrections destroy some of the
symmetries of the Born terms. However,
there are several symmetry relations which remain valid even at
one loop:
\beqn
\label{symmetry1}
a^{(1)}_{Q\Qb}(-,-,+) &=& - a^{(1)}_{Q\Qb} (+,+,+) |_{\qb \lra q ,\Qb \lra Q},
\\ \label{symmetry2}
\vspaceinarray
a^{(1)}_{Q\Qb}(+,-,+) &=& -a^{(1)}_{Q\Qb} (-,+,+) |_{\qb \lra q ,\Qb \lra Q},
\\ \label{symmetry3}
\vspaceinarray
a^{(1)}_{Q\qb}(-,-,+)&=&- a^{(1)}_{Q\qb} (+,+,+) |_{\qb \lra Q ,q \lra \Qb}.
\eeqn
We find that $\S_{ij}$, $\R_{ij}$, $\E_{ij}$ and $\C_{Q\Qb}$ respect
these symmetries separately. As a result the $\D_{Q\Qb}$ terms must
also satisfy the relations (\ref{symmetry1}) and (\ref{symmetry2}).
In the case of $\C_{Q\qb}$, if we factor the helicity dependence into
the Born terms as in eq.\ (\ref{CQqb}),  then the symmetry relation
(\ref{symmetry3}) is slightly violated which is compensated by a
corresponding change in $\D_{Q\qb}$:
\beq
\D_{Q\qb}(-,-,+) = - \D_{Q\qb} (+,+,+) |_{\qb \lra Q ,q \lra \Qb}
 + \frac{2}{3} (\Nc-\Nf) a^{(0)}_{Q \qb} (-,-,+)
    \ln \left( \frac{\s \Qb Q}{s \qb q} \right).
\eeq
In order to give the results in a more compact form, we
introduce the following notation\footnote{Note that the function
$\L n{^{i j}_{k l}}$ has $-2n$ mass dimensions. These functions
are related to similar ones introduced in ref.\ \cite{Ber93}.}:
\beqn
\L0{^{i j}_{k l}} &=& \ln \left(\frac{- \s i j}{ \mu^2}\right) -
                 \ln \left(\frac{- \s k l}{ \mu^2}\right), \qquad
\qquad
\L1{^{i j}_{k l}} = \frac{1}{\s k l - \s i j} \L0{^{i j}_{k l}},  \\
\L2{^{i j}_{k l}} &=&  \frac{1}{(\s k l - \s i j)^2}
                \L0{^{i j}_{k l}} + \frac{1}{\s k l (\s k l - \s i j)} , \;
\L3{^{i j}_{k l}} =  \frac{1}{(\s k l - \s i j)^3}
                     \L0{^{i j}_{k l}} +
   \frac{(\s i j + \s k l)}{2\ \s i j \s k l (\s k l - \s i j)^2}.
\nn \eeqn
Now the $\D$ terms take the form

\beqn
\lefteqn{-\i \D_{Q \Qb} (+,+,+) = } \\
&& \Nc \left(\frac{\A \qb Q^2 \A \Qb g \B Q g}{ \A \qb q \A Q g^2 } +
   \frac{3\ \A \qb \Qb \A \qb Q \B Q g }{2\ \A \qb q \A Q g} \right)
      \L1{^{ \Qb g}_{\qb q}} \nn \\
&-&
 \frac{\Nc^2+1}{\Nc} \left[
   \frac{\A \qb \Qb \A \Qb Q \B q Q}{2\ \A \Qb g \A Q g \s \qb q} +
   \frac{\A \qb \Qb \B q g}{\A Q g \s \qb q}  +
   \frac{\A \qb Q \A \Qb Q \B q Q \B Q g}{2\ \A Q g }
      \L2{^{ \Qb g}_{\qb q}} \right. \nn \\
&& \quad \left.  -
   \frac{\A \qb \Qb \A \qb Q}{\A \qb q \A Q g^2}
     \L0{^{ \Qb Q}_{\qb q}} +
   \frac{\A \qb Q \A \Qb g \B q g }{ \A Q g^2}
     \L1{^{ \Qb Q}_{\qb q}} -
   \frac{\A \qb g \A \Qb Q \B q g \B Q g}{ \A Q g}
     \L2{^{ \Qb Q}_{\qb q}} \right] \nn \\
&-& \frac{1}{\Nc}
\left[
   \frac{\A \qb \Qb}{\A q g \A Q g}
      \L0{^{ \Qb g}_{\qb Q}} +
   \frac{\A \qb \Qb \A \qb Q }{\A \qb q \A Q g^2}
      \L0{^{ \Qb g}_{\qb q}} +
   \frac{3\ \A \qb \Qb^2}{2\ \A \qb q \A \Qb g \A Q g}
      \L0{^{ \Qb Q}_{\Qb g}}  \right. \nn \\
&& \quad -  \left(
\frac{\A \qb Q \A q g \A \Qb Q^2 \B q Q }{ \A q Q \A \Qb g \A Q g^2}-
   \frac{\A \qb \Qb \A \Qb Q \B q Q}{2\ \A \Qb g \A Q g} \right)
      \L1{^{ \Qb g}_{\qb q}} \nn \\
&& \quad + \left.
   \frac{\A \qb g \A q \Qb \B q g}{\A q g \A Q g}
      \L1{^{ q \Qb}_{\qb Q}}  -
   \frac{\A \qb Q \A q \Qb^2 \B q Q}{\A q Q \A q g \A \Qb g}
      \L1{^{ \Qb g}_{\qb Q}} \right], \nn
\eeqn

\beqn
\lefteqn{-\i \D_{Q \Qb} (-,+,+) = } \\
&&  \Nc \left[
  \frac{3\ \A q \Qb^2}{2\ \A \qb q \A \Qb g \A Q g}
     \L0{^{ \Qb g}_{\qb q}} +
  \frac{3\  \A q \Qb \A \Qb Q \B \qb Q }{ \A \Qb g \A Q g}
     \L1{^{ \Qb g}_{\qb q}} \right]
 +\frac{1}{\Nc}  \frac{3\ \A q \Qb^2}{2\ \A \qb q \A \Qb g \A Q g}
     \L0{^{ \Qb Q}_{\Qb g}} \nn \\
&-&
 \frac{\Nc^2+1}{\Nc}  \left[
 \frac{\A q \Qb \B \qb g}{ \A Q g \s \qb q} +
 \frac{\A q \Qb }{\A \qb g \A Q g}
     \L0{^{ q Q}_{\Qb g}}+
 \frac{\A q \Qb \A q Q}{\A \qb q \A Q g^2}
     \L0{^{ \Qb Q}_{\Qb g}} \right. \nn \\
& & \quad  \quad -
 \frac{\A \qb \Qb \A q g \B \qb g}{\A \qb g \A Q g }
     \L1{^{ q Q}_{\qb \Qb}} +
 \frac{\A q Q \A \Qb g \B \qb g}{\A Q g^2}
     \L1{^{ \Qb Q}_{\qb q}} -
 \frac{2\ \A q g \A \Qb Q^2 \B \qb Q}{\A \Qb g \A Q g^2}
     \L1{^{ \Qb g}_{\qb q}} \nn \\
& & \quad \quad +
\frac{\A \qb g \A q Q \A \Qb Q^2 \B \qb Q}{\A \qb Q \A \Qb g \A Q g^2}
     \L1{^{ \Qb g}_{\qb q}} +
 \frac{\A \qb \Qb^2 \A q Q \B \qb Q}{\A \qb Q \A \qb g \A \Qb g}
     \L1{^{ \Qb g}_{q Q}} \nn \\
& & \quad  \quad \left. -
 \frac{\A q g \A \Qb Q \B \qb g \B Q g}{\A Q g}
     \L2{^{ \Qb Q}_{\qb q}} -
 \frac{\A \qb q \A \Qb Q^2 \B \qb Q^2}{2\ \A \Qb g \A Q g}
     \L2{^{ \Qb g}_{\qb q}}  \right], \nn
\eeqn

\beqn
\lefteqn{-\i \D_{Q \qb} (+,+,+) =} \\
&& \hspace*{-1.3em} \frac{2 (\Nc \!- \! \Nf)}{3} \! \left[
   \frac{\A \qb \Qb^2 \B \Qb g }{\A \qb q \A Q g}
        \L1{^{ \Qb Q}_{\qb q}} \! -
   \frac{\A \qb \Qb \B q g \B Q g }{2}
        \L2{^{ \Qb Q}_{\qb q}}  \!  +
   \A \qb g \A \Qb Q \B q g \B Q g^2
        \L3{^{ \Qb Q}_{\qb q}} \right]  \nn \\
&+& \frac{ \Nc^2 +1}{\Nc}  \left[
  - \frac{\A \qb \Qb^2 \A q Q}{2\ \A \qb q \A q g \A \Qb Q \A Q g} -
   \frac{\A \qb \Qb \B q g}{2\ \A Q g \s \qb q } -
   \frac{\A \qb q \A \Qb g \A \qb \Qb}{\A q g^2 \A \Qb Q \A \qb g}
       \L0{^{ \Qb Q}_{\qb g}} \right. \nn \\
&& \qquad +
   \frac{\A \qb q^2 \B q Q (\A \Qb q \A Q g + \A \Qb Q \A q g )}
        {\A \qb g \A q Q \A q g^2} \L1{^{ \Qb Q}_{\qb g}}
  -\frac{2\ \A \qb Q^2 \A q \Qb \B q Q}{\A \qb g \A q Q \A Q g }
       \L1{^{ q \Qb}_{\qb g}} \nn \\
& & \qquad + \left(
  -\frac{3\ \A \qb \Qb \B q g}{2\ \A Q g}
  +\frac{\A \qb g \A q \Qb \B Q g}{\A q g^2}
  -\frac{2\ \A \qb Q \A \Qb g \B Q g}{\A q g \A Q g} \right)
       \L1{^{ \Qb Q}_{\qb q}} \nn \\
&& \qquad \left. +\frac{\A \Qb Q}{2} \left(
   \frac{\A \qb g \B q g \B Q g }{\A Q g}\L2{^{ \Qb Q}_{\qb q}}
  -\frac{\A \qb g \B Q g ^2}{\A q g}\L2{^{ \qb q}_{ \Qb Q}}
  -\frac{\A \qb q^2 \B q Q^2}{\A \qb g \A q g}
       \L2{^{ \qb g}_{\Qb Q}} \right)  \right] \nn \\
&+& \Nc \left[
   \frac{\A \qb \Qb \A \qb Q \A \Qb g}{\A \qb g \A q g \A \Qb Q \A Q g}
       \L0{^{ \Qb Q}_{q \Qb}}
  +\left( \frac{\A \qb g \A q \Qb}{\A q g \A \qb \Qb} + \frac{1}{2}\right)
   \frac{\A \qb \Qb^2 \A q Q}{\A \qb q \A q g \A \Qb Q \A Q g}
   \L0{^{ \Qb Q}_{\qb q}}  \right. \nn \\
&& \qquad +
 \frac{3\ \A \qb \Qb^2}{2\ \A \qb g \A q g \A \Qb Q}
       \L0{^{ \Qb Q}_{\qb g}} +
   \left( \frac{\A q \Qb \A Q g}{\A q g \A \Qb Q} - 2 \right)
   \frac{\A \qb \Qb \A \qb Q}{\A \qb g \A q Q \A Q g}
       \L0{^{ q \Qb}_{\qb g}} \nn \\
&& \qquad \left. -3\ \A \qb Q \left(
   \frac{\A \Qb Q \B Q g}{\A q Q \A Q g}
       \L1{^{ q \Qb}_{\qb g}}
  +\frac{\A \qb g \A q \Qb \B Q g}{\A \qb q \A q g \A Q g}
       \L1{^{ \Qb Q}_{\qb q}}
  -\frac{\A \qb q \A q \Qb \B q Q}{\A \qb g \A q Q \A q g}
       \L1{^{ \Qb Q}_{\qb g}} \right) \right] \nn \\
&+& \frac{1}{\Nc} \left[
     \frac{\A \qb q \A \qb \Qb \A \Qb g}{\A \qb g \A q g^2 \A \Qb Q}
        \L0{^{ \Qb Q}_{\qb q}} +
     \frac{\A \qb \Qb^2}{2\ \A \qb g \A q g \A \Qb Q}
        \L0{^{ \qb g}_{\qb q}} +
     \frac{2\ \A \qb \Qb}{\A q g \A Q g}
        \L0{^{ \qb g}_{\qb q}} \right], \nn
\eeqn

\beqn
\lefteqn{-\i \D_{Q \qb} (-,+,+) = } \\
&&  \frac{\Nc^2 + 1}{\Nc} \left[
   \frac{ \A \qb Q \A q \Qb^2}{2\ \A \qb q \A \qb g \A \Qb Q \A Q g} -
   \frac{\A q \Qb \B \qb g}{2\ \A Q g \s \qb q} +
   \frac{\A q g \B Q g^2}{2\ \A \qb g \B \Qb Q \s \qb q} \right. \nn \\
&& \qquad \left. -
   \frac{\A q \Qb \B Q g}{\A \qb g}
       \L1{^{ \Qb Q}_{\qb q}} +
   \frac{\A q g \A \Qb Q \B Q g (\s Q g -\s \qb g)}{2\ \A \qb g \A Q g}
       \L2{^{ \Qb Q}_{\qb q}} \right] \nn \\
&+& \frac{\Nf}{3} \left[
 -    \frac{2\ \A q \Qb^2 \A q Q}{\A \qb q \A q g \A \Qb Q \A Q g}
       \L0{^{ \Qb Q}_{\qb q}} -
   \frac{2\ \A q \Qb^2 \B \qb g}{ \A q g \A \Qb Q}
       \L1{^{ \Qb Q}_{\qb q}} \right. \nn \\
&& \quad \left. +
   \A q \Qb \B \qb g \B Q g
       \L2{^{ \qb q}_{\Qb Q}} -
   2\ \A \qb q \A \Qb g \B \qb g^2 \B Q g
       \L3{^{ \Qb Q}_{\qb q}} \right] \nn \\
&-& \frac{1}{\Nc} \left[
   \frac{2\ \A q \Qb }{\A \qb g \A Q g}
        \L0{^{ \qb \Qb}_{\Qb Q}} -
   \frac{2\ \A q Q \A \Qb g \B Q g }{\A \qb g \A Q g}
        \L1{^{ q Q}_{\qb \Qb}} \right. \nn \\
&& \quad \left.
   -\left(\frac{3\ \A \Qb q }{2\ \A \Qb g } +
          \frac{2\ \A \qb q }{\A \qb g} \right)
    \frac{\A \Qb g \B \qb g}{\ \A Q g} \L1{^{ \Qb Q}_{\qb q}} \right] \nn \\
&+& \Nc \left[
   -\frac{3\ \A q \Qb^2 \A \qb \Qb}{2\ \A \qb q \A \Qb Q \A \qb g \A \Qb g}
     \L0{^{ \Qb Q}_{\qb q}}
   +\left(
   \frac{3\ \A q \Qb \A q g \B Q g (\A \qb Q \A \Qb g + \A \qb \Qb \A Q g )}
        {2\ \A \qb q \A \qb g \A \Qb g \A Q g} \right. \right. \nn \\
&& \qquad \left. \left.
   -\frac{2\ \A q \Qb^2 \B \Qb g}{3\ \A \qb q \A Q g} \right)
        \L1{^{ \Qb Q}_{\qb q}}
   -\frac{\A q \Qb \B \qb g \B Q g }{3}
        \L2{^{ \Qb Q}_{\qb q}} +
   \frac{2\ \A q g \A \Qb Q \B \qb g \B Q g^2 }{3}
        \L3{^{ \Qb Q}_{\qb q}} \right], \nn
\eeqn

\beqn
\lefteqn{-\i \D_{Q \qb} (+,-,+) = } \\
&& \hspace*{-0.9cm} \frac{2 (\Nf \!- \! \Nc)}{3} \! \left[
   \frac{\A \qb Q^2 \B \Qb g}{\A \qb q \A Q g }
       \L1{^{ \Qb Q}_{\qb q}} +
   \frac{\A \qb Q \B q g \B \Qb g}{2}
       \L2{^{ \Qb Q}_{\qb q}} +
   \A \qb g \A \Qb Q \B q g \B \Qb g^2
       \L3{^{ \Qb Q}_{\qb q}}\right] \nn \\
&+& \Nc \left[
   \frac{\A \qb Q^3}{2\ \A \qb q \A \qb g \A \Qb Q \A Q g}  -
   \left( \frac{\A \qb g \A q Q^2 \B q g}{\A q g^2 \A \Qb Q } +
   \frac{3\ \A \qb Q \A q Q \B q g}{2\ \A q g \A \Qb Q} \right)
        \L1{^{ \Qb Q}_{\qb g}} \right.
\nn \\
&& \quad -
   \left( \frac{\A \qb g \A q Q^2 \B q g}{ \A q g^2 \A \Qb Q }  +
   \frac{\A \qb Q \A q \Qb \A Q g \B q g }
       {2\ \A q g \A \Qb Q \A \Qb g }  -
   \frac{ \A \qb Q^2  \B Q g}
        {2\ \A \qb q \A \Qb g } +
   \frac{\A \qb \Qb^2 \A Q g  \B \Qb g}
        {\A \qb q \A \Qb g^2} \right) \L1{^{ \Qb Q}_{\qb q}} \nn \\
&& \quad -
   \left( \frac{3\ \A \qb \Qb \A \qb Q \B \Qb g}{2\ \A \qb q \A \Qb g } +
   \frac{\A \qb \Qb^2 \A Q g \B \Qb g}{ \A \qb q \A \Qb g^2} \right)
      \L1{^{ Q g}_{\qb q}} -
   \frac{\A \qb \Qb \A \Qb Q \B q \Qb \B \Qb g}{2\ \A \Qb g }
        \L2{^{ Q g}_{\qb q}} \nn \\
&& \quad + \left.
   \frac{\A \qb g \A \Qb Q \B q g \B \Qb g}{2\ \A \Qb g }
       \L2{^{ \Qb Q}_{\qb q}} +
   \frac{\A \qb q \A Q g \B q g \B \Qb g}{2\ \A q g }
        \L2{^{\qb q}_{\Qb Q}} -
   \frac{\A \qb q \A q Q \B q \Qb \B q g}{2\ \A q g }
        \L2{^{ \qb g}_{\Qb Q}} \right] \nn \\
&+& \frac{1}{\Nc} \left[
   \frac{\A \qb Q^2 \A q Q}{2\ \A \qb q \A q g \A \Qb Q \A Q g} +
   \frac{\A \qb Q \A \Qb Q \B q \Qb}{2\ \A \Qb g \A Q g \s \qb q} +
   \frac{\A \qb g \A \Qb Q \B \Qb g}{2\  \A q g \A \Qb g \s \qb q} +
   \frac{\A \qb g \B \Qb g^2}{2\ \A q g \B \Qb Q \s \qb q} \right. \nn \\
&& \quad
   -\frac{2\ \A \qb Q \A q Q}{\A q \Qb \A q g \A Q g}
      \L0{^{ \qb \Qb}_{\qb q}}
  -\frac{\A \qb Q^2}{2\ \A \qb g \A q g \A \Qb Q}
      \L0{^{ \qb g}_{\qb q}} -
   \frac{2\ \A \qb q \A \qb Q }{\A \qb g \A q \Qb \A q g}
      \L0{^{ q Q}_{\qb g}} \nn \\
&& \quad
  +\left(\frac{\A \qb g \A \Qb Q \A \qb Q }{\A \Qb g^2 \A Q g \A \qb q}
  -\frac{\A \qb Q \A \qb q \A Q g}{\A \qb g \A q g^2 \A \Qb Q} \right)
    \L0{^{ \Qb Q}_{\qb q}}
  +\frac{\A \qb q \A Q g \B \Qb g}{\A q g^2 \s \Qb Q}
      \L0{^{ \Qb Q}_{\qb g}} \nn \\
&& \quad + \left(
   \frac{2\ \A \qb Q \A \Qb Q}{\A q \Qb \A \Qb g \A Q g} -
   \frac{\A \qb g \A \Qb Q \B q g}{\A \Qb g^2 \s \qb q} -
   \frac{\A \qb Q^2 }{2\ \A \qb q \A \Qb g \A Q g} \right)
     \L0{^{ Q g}_{\qb q}} \nn \\
&& \quad - \left(
   \frac{\A \qb g \A \Qb Q \B q g}{\A \Qb g^2} +
   \frac{2\ (\A \qb Q \A \Qb g + \A \qb g \A \Qb Q)\B \Qb g}{\A q g \A \Qb g}
  +\frac{\A \qb g^2 \A q Q \A \Qb Q \B \Qb g}
        {2\ \A \qb q \A q g \A \Qb g \A Q g} \right. \nn \\
&& \qquad \left.
   -\frac{\A \qb g \A q Q \B \Qb g}{\A q g^2} \right)
      \L1{^{ \Qb Q}_{\qb q}}
   -\frac{2\ \A \qb \Qb \A Q g \B \Qb g}{\A q g \A \Qb g }
      \L1{^{ q Q}_{\qb \Qb}} +
   \frac{2\ \A \qb \Qb \A \Qb Q \B \Qb g}{\A q \Qb \A \Qb g }
      \L1{^{ q Q}_{\qb g}} \nn \\
&& \quad +\left(
   \frac{2\ \A \qb q^2 \A \Qb Q \B q \Qb }{\A \qb g \A q \Qb \A q g }+
   \frac{\A \qb q^2 \A Q g \B \qb g \B q \Qb}{\A q g^2  \s \Qb Q} \right)
      \L1{^{ \Qb Q}_{\qb g}} +
   \frac{2\ \A \qb q \A q Q \B q g}{\A q \Qb \A q g}
      \L1{^{ Q g}_{\qb \Qb}}  \nn \\
&& \quad  + \left(
   \frac{2\ \A \qb q \A \Qb Q^2 \B q \Qb}{\A q \Qb \A \Qb g \A Q g}+
   \frac{\A \qb g \A \Qb Q^2 \B q \Qb \B Q g}{ \A \Qb g^2 \s \qb q}  \right)
      \L1{^{ Q g}_{\qb q}} +
 \frac{\A \qb q^2 \A \Qb Q \B q \Qb^2}{2\ \A \qb g \A q g}
      \L2{^{ \qb g}_{\Qb Q}}  \nn \\
&& \quad \left. +
  \frac{\A \qb g \A \Qb Q \B \Qb g ( \s q Q -\s \qb \Qb)}{2\ \A q g \A \Qb g}
 \L2{^{ \Qb Q}_{\qb q}} +
  \frac{\A \qb q \A \Qb Q^2 \B q \Qb^2}{2\ \A \Qb g \A Q g}
      \L2{^{ Q g}_{\qb q}} \right]. \nn
\eeqn

The $\E$ contributions can conveniently be given in terms of some
auxiliary functions defined as follows.
\begin{equation}
\F (i,j,k) =
-\Li \left(1-{s_{mn}\over s_{ij}}\right)
-\Li \left(1-{s_{mn}\over s_{jk}}\right)
-\frac{1}{2}\ln^2\left({s_{ij}\over s_{jk}}\right)-{\pi^2\over 6},
\end{equation}
where $\Li$ is the dilogarithm and $m,n$ denote the complementary
labels to $(i,j,k)$ in the set of labels  $(\qb,\Qb,Q,q,g)$
\beq
\{m,n\}=\{\qb,\Qb,Q,q,g\}\backslash \{i,j,k\}.
\eeq
We also use  the functions
\beq
h(i,j,k,l)={ \la ik \ra ^2 \la jl \ra ^2\over  \la il \ra ^2 \la jk \ra ^2},
\eeq
\beq
\tilde{h}(i,j,k,l)=h(i,j,k,l)^{((1-\delta_{ik})(1-\delta_{jl}))}
\eeq
and
\beqn
\G(r,a,A,g,R)&=&h(r,a,g,R)^{\delta_{RA}} \times \F(a,A,g),\\
\G(r,g,a,A,R)&=&h(r,g,A,R)^{\delta_{ra}} \times \F(g,a,A),\nn
\eeqn
where  $a \in \{ q,\qb\}$ and $A \in \{Q,\Qb\}$ respectively.
Then the $ \E_{Q\qb}(h_q,h_Q)$ term for arbitrary $h_q,h_Q$ quark
helicity configuration is given as
\beqn
\E_{Q\qb}(h_q,h_Q,+) &=& \\ \nn
-a^{(0)}_{Q\qb}(h_q,h_Q,+) &&  \hspace*{-1  cm} \left\{ \Nc\left[
\F(\qb,g,Q) + h(r,\qb,\Qb,R) \F(\qb,q,\Qb)
+ h(r,q,g,R) \F(q,\qb,g) \right.\right. \\ \nn
&&+ \left.  h(r,g,\Qb,R) \F(g,Q,\Qb)
 + h(r,q,Q,R) \F(q,\Qb,Q)\right] \\ \nn
&& \hspace*{-1  cm}- {1\over \Nc}\left[
-h(r,q,\Qb,R)\left(\F(q,Q,\Qb)+\F(q,\qb,\Qb)\right)\right.
+\F(q,g,\Qb)\\ \nn
&&-h(r,\qb,Q,R)\left(\F(\qb,\Qb,Q)+\F(\qb,q,Q)\right)+\F(\qb,g,Q) \\
\nn
&&+h(r,\qb,\Qb,R)\left(\F(\qb,Q,\Qb)+\F(\qb,q,\Qb)\right)-\F(\qb,g,\Qb)\\
\nn
&&+ h(r,q,Q,R) \left.\left.
\left( \F(q,\qb,Q)+\F(q,\Qb,Q) \right) -\F(q,g,Q) \right] \right\} \\ \nn
+ a^{(0)}_{Q\Qb}(h_q,h_Q,+) && \hspace*{-1  cm}
  \left\{\Nc h(r,g,\Qb,R) \F(g,Q,\Qb) \right.\\ \nn
&& \hspace*{-1  cm} - {1\over\Nc}\left[-\F(\qb,g,\Qb)+h(r,\qb,g,R)
\F(\qb,q,g)\right.  \\ \nn
&&+\F(\Qb,g,q)-h(r,q,g,R) \F(q,\qb,g)
-\tilde{h}(r,\Qb,g,R) \F(\Qb,Q,g) \\ \nn
&&-h(r,q,\Qb,R)\left(\F(q,Q,\Qb)-\F(q,\qb,\Qb)\right) \\ \nn
&&+h(r,\qb,\Qb,R)\left(\F(\qb,Q,\Qb)-\F(\qb,q,\Qb)\right) \\ \nn
&&+\G(r,\qb,\Qb,g,R)-\G(r,\qb,Q,g,R) \\ \nn
&&+\left.\left.\G(r,q,Q,g,R)-\G(r,q,\Qb,g,R) \right]\right\} \\ \nn
+a^{(0)}_{q\qb}(h_q,h_Q,+) && \hspace*{-1  cm}
   \left\{\Nc h(r,q,g,R) \F(q,\qb,g) \right.\\ \nn
&& \hspace*{-1  cm}
-{1\over\Nc}\left[\F(\Qb,g,q)-h(r,g,\Qb,R)\F(g,Q,\Qb)
\right.  \\ \nn
&&-\F(Q,g,q)+h(r,g,Q,R) \F(g,\Qb,Q)
-\tilde{h}(r,g,q,R) \F(g,\qb,q)\\ \nn
&&-h(r,q,\Qb,R)\left(\F(q,\qb,\Qb)-\F(q,Q,\Qb)\right) \\ \nn
&&+h(r,q,Q,R)\left(\F(q,\qb,Q)-\F(q,\Qb,Q)\right) \\ \nn
&&+\G(r,g,\qb,\Qb,R)-\G(r,g,q,\Qb,R) \\ \nn
&&+\left.\left.\G(r,g,q,Q,R)-\G(r,g,\qb,Q,R) \right]\right\}. \\ \nn
\eeqn
As noted before, the $\E_{Q\Qb}$ terms satisfy the symmetry relations
(\ref{symmetry1}) and (\ref{symmetry2}), therefore it is sufficient
to present the $\E_{Q\Qb}$ type contributions for $h_q=+$ and arbitrary
$h_Q$:
\beqn
&& \E_{Q\Qb}(+,h_Q,+)=
a^{(0)}_{Q\Qb}(+,h_Q,+)\,\H_{Q\Qb}^{h_Q}(\qb,\Qb,Q,q,g)\\ \nn
&&+a^{(0)}_{Q\qb}(+,h_Q,+)\,\H_{Q\qb}^{h_Q}(\qb,\Qb,Q,q,g)
+a^{(0)}_{q\Qb}(+,h_Q,+)\,\H_{q \Qb}^{h_Q}(\qb,\Qb,Q,q,g),
\eeqn
where the auxiliary functions, $\H_{ij}^{h_Q}$ have the form
\beqn
\lefteqn{\H_{Q \Qb}^- (\qb,\Qb,Q,q,g) = } \\
&&\Nc \left[ \F(\qb,g,q) +\F(Q,q,g) - \F(\Qb,g,\qb) -
\F(Q,g,q)  -\F(Q,q,\qb) - \F(g,q,\qb) \right.  \nn \\
&&\quad + h(\qb,g,\Qb,Q) \F(g,\qb,\Qb)  \left.- h(\qb,q,\Qb,Q)
\F(q,\qb,\Qb) -  h(\qb,g,q,Q) \F(g,\qb,q) \right] \nn \\
&&\hspace*{-1em}-\frac{1}{\Nc} \left[
\F(\qb,q,Q) - \F(\qb,q,\Qb) - \F(Q,\qb,q) + \F(\qb,\Qb,g) -\F(Q,\Qb,g)
\right.  \nn \\
&&\quad - \F(\qb,Q,g) - \F(q,\Qb,g) + h(\qb,q,\Qb,Q) \F(q,\qb,\Qb) \nn \\
&&\quad\left. +
h(\qb,q,g,Q)\F(q,Q,g)-h(\qb,\Qb,g,Q)\F(\Qb,Q,g)\right], \nn \\
\lefteqn{\H_{Q \Qb}^+  (\qb,\Qb,Q,q,g) = }\\
&&\Nc \left[ \F(\qb,g,q) +\F(Q,q,g) - \F(\Qb,g,\qb) -
\F(Q,g,q)  -\F(Q,q,\qb) - \F(g,q,\qb) \right.  \nn \\
&&\quad +\F(g,\qb,\Qb)\left.
-\F(q,\qb,\Qb)-h(\qb,g,q,\Qb)\F(g,\qb,q)\right] \nn \\
&&\hspace*{-1em}-\frac{1}{\Nc} \left[
\F(\qb,q,Q) - \F(\qb,q,\Qb) + \F(\Qb,\qb,q) + \F(\qb,\Qb,g) \right. \nn \\
&&\quad +  \F(q,Q,g) - \F(\qb,Q,g) -\F(\Qb,Q,g)  \nn \\
&&\quad\left. - h(\qb,q,Q,\Qb) \F(Q,\qb,q) - h(\qb,q,g,\Qb)
\F(q,\Qb,g) - h(\qb,Q,g,\Qb) \F(Q,\Qb,g) \right], \nn \\
\lefteqn{ \H_{Q \qb}^{h_Q}  (\qb,\Qb,Q,q,g) = }\\
&&\Nc \left[ \F(\qb,g,\Qb) - \F(\qb,g,q) +
\F(Q,q,\qb) - \F(Q,q,g) - \F(Q,\Qb,\qb) \right. \nn \\
&&\quad \left. - h(\qb,g,\Qb,R) \F(g,\qb,\Qb) + h(\qb,g,q,R)
\F(g,\qb,q) + h(\qb,g,Q,R) \F(g,\Qb,Q) \right], \nn \\
\lefteqn{ \H_{q \Qb}^{h_Q}  (\qb,\Qb,Q,q,g) = } \\
&&\Nc \left[
\F(q,g,Q) - \F(q,g,\qb) - \F(Q,q,g) + \F(\qb,q,g) \right. \nn \\
&&\quad + h(\qb,q,\Qb,R) \F(q,\qb,\Qb) - h(\qb,g,\Qb,R) \F(g,\qb,\Qb)
+ h(\qb,g,\Qb,R) \F(g,Q,\Qb) \nn \\
&&\quad \left. - h(\qb,q,\Qb,R) \F(q,Q,\Qb) + h(\qb,g,q,R) \F(g,\qb,q)
- h(\qb,q,g,R) \F(q,\qb,g) \right]. \nn
\eeqn

The above results are valid in the unphysical region, where the dot
products are negative, therefore the arguments of the logarithms and
dilogarithms is away from the branch cuts. To obtain the amplitudes in
any physical channel, one has to continue analytically to the
corresponding physical region and make the usual substitution
\beq
s_{ij} \to s_{ij} + \i \eta.
\eeq
This defines all functions in a unique way.

\def\np#1#2#3  {{\it Nucl. Phys. }{\bf #1} (19#3) #2}
\def\nc#1#2#3  {{\it Nuovo. Cim. }{\bf #1} (19#3) #2}
\def\pl#1#2#3  {{\it Phys. Lett. }{\bf #1} (19#3) #2}
\def\pr#1#2#3  {{\it Phys. Rev. }{\bf #1} (19#3) #2}
\def\prl#1#2#3  {{\it Phys. Rev. Lett.}{\bf #1} (19#3) #2}
\def\prep#1#2#3{{\it Phys. Rep. }{\bf #1} (19#3) #2}

\end{document}